\documentclass[prl,superscriptaddress,footinbib,twocolumn,floatfix]{revtex4}
\usepackage{graphics,epsfig,color,subfigure}
\usepackage{amssymb}

\makeatletter
\renewcommand{\@makecaption}[2]{
  \vskip\abovecaptionskip
  \sbox\@tempboxa{\small\sf #1: #2}%
  \ifdim \wd\@tempboxa >\hsize
  \small\sf #1: #2\par
  \else
    \global \@minipagefalse
    \hb@xt@\hsize{\hfil\box\@tempboxa\hfil}%
  \fi
  \vskip\belowcaptionskip}
\makeatother

\def\be{\begin{eqnarray}}
\def\ee{\end{eqnarray}}

\newcommand{\eqn}[1]{(\ref{#1})}

\def\Dslash{\,\,{\raise.15ex\hbox{/}\mkern-12mu D}}
\def\Dbarslash{\,\,{\raise.15ex\hbox{/}\mkern-12mu {\bar D}}}
\def\delslash{\,\,{\raise.15ex\hbox{/}\mkern-9mu \partial}}
\def\delbarslash{\,\,{\raise.15ex\hbox{/}\mkern-9mu {\bar\partial}}}
\def\pslash{\,\,{\raise.15ex\hbox{/}\mkern-9mu p}}
\def\calDslash{\,\,{\raise.15ex\hbox{/}\mkern-12mu {\cal D}}}

\def\lae{\mathrel{\mathop{\smash{\lower .5 ex \hbox{$\stackrel<\sim$}}}}}
\def\lae{\mathrel{\mathop{\smash{\lower .5 ex \hbox{$\stackrel>\sim$}}}}}

\begin{document}

\title{Quantum Critical Transport and the Hall Angle}

\author{Mike Blake}
\author{Aristomenis Donos}

\affiliation{Department of Applied Mathematics and Theoretical Physics,
University of Cambridge, CB3 0WA, UK}

\begin{abstract}
In this letter we study the Hall conductivity in holographic models where translational invariance is broken by a lattice. We show that generic holographic theories will display a different temperature dependence in the Hall angle as to the DC conductivity. Our results suggest a general mechanism for obtaining an anomalous scaling of the Hall angle in strongly interacting quantum critical systems. 
\end{abstract}


\maketitle










\paragraph{Introduction}
\label{Intro}

Many interesting experimental systems are governed by quantum criticality \cite{qpt}. In particular, it has long been suggested that the anomalous scaling seen in the transport properties of the strange metal phase is indicative of quantum critical behaviour \cite{damle, vandermarel, phillips, zaanenrev}. 
Unfortunately, performing explicit calculations of strongly interacting transport is an extremely challenging task.  As a result, many of the  experimental anomalies remain as mysterious as ever.

Particularly puzzling is the scaling of the Hall angle. Simple models of transport, such as the Drude model, suggest that the Hall angle, $\theta_H$, should be proportional to the DC conductivity, $\sigma_{DC}$. However, experimental measurements of the Hall angle in the cuprates show a Fermi liquid behaviour $\theta_H \sim 1/T^2$, which is hard to reconcile with the ubiquitous linear resistivity of strange metals \cite{ong}. 


Previous theoretical attempts to explain this behaviour begin by proposing that transport in these materials is governed by two different relaxation times. In \cite{Anderson} it is suggested that spinons and holons scatter at different rates, whilst in \cite{Coleman} it is charge conjugation odd and even quasiparticles that behave independently. 

In this letter, we use holography to study the Hall angle in strongly interacting critical theories. In particular we obtain simple analytic expressions for the conductivity tensor of a large class of holographic lattice models in the presence of a magnetic field. Our calculations indicate that within these holographic models the Hall angle will generically display a different temperature dependence to the DC conductivity.

Further, these results suggest a general mechanism for obtaining an anomalous scaling in the Hall angle that can apply beyond these holographic models. The essential feature is the existence of two distinct terms that are additive, following an `inverse-Matthiessen' law, in the conductivity. Firstly, there is a `charge-conjugation symmetric' contribution, denoted by $\sigma_{ccs}$, which is analogous to that found at relativistic quantum critical points.  In addition there is a contribution to the current from any net charge density. This current is relaxed by the effects of momentum dissipation and results in a second term in the conductivity, $\sigma_{diss}$. 

In contrast, we find that there is only a single term that controls the scaling of the Hall angle, which is proportional to $\sigma_{diss}$. In particular, there is no additive contribution to $\theta_H$ analogous to the `charge-conjugation symmetric' term in $\sigma_{DC}$. When momentum relaxation is strong, we therefore find that the Hall angle and DC conductivity can be very different. In this regime, the `charge-conjugation symmetric' contribution dominates the conductivity, $\sigma_{DC} \sim \sigma_{ccs}$, whilst the Hall angle continues to scale as  $\theta_H \sim \sigma_{diss}$. 

Although we have used specific holographic models to identify this mechanism, our arguments can be applied more generally - in particular, we note, to the results of relativistic magnetohydrodynamics studied in \cite{hydro}. As such, we end this letter with a brief discussion of how these ideas could be applied to the experimental phenomenology of the cuprates. 



\paragraph{Holographic Models}

The holographic correspondence allows us to study certain strongly coupled systems in terms of gravitational physics in anti-de Sitter space. Our goal in this section is to study the Hall angle in a large class of typical holographic models. To this end we begin by considering an Einstein-Maxwell-Dilaton theory, with additional axion fields which we shall use to break translational invariance in the boundary theory.
\be
S = \int \mathrm{d}^4x \sqrt{-g} \bigg [ R - \frac{1}{2}( \partial \phi)^2  + \Phi(\phi) ((\partial {\chi_1})^2  + (\partial {\chi_2})^2) \nonumber \\ + V(\phi)  - \frac{Z(\phi)}{4} F^2 \bigg]
\nonumber
\label{action}
\ee
These models, originally introduced in \cite{Qlattices}, can be viewed as the polar decomposition of a theory with two identical complex
scalar fields which are assumed to have the same magnitude $\phi_1 = \phi e^{i \chi_1}$, $\phi_2 = \phi e^{i \chi_2}$. 
If we make the identification $\Phi(\phi) = \phi^2$, this corresponds to the original scalars $\phi_i$ living in the complex plane. Other forms, in particular 
exponentials, can be obtained from scalars taking values in e.g. hyperbolic space. 
In addition we have allowed for an arbitrary potential for the magnitude of these lattices, and a coupling to the electromagnetic field via
a dilaton-like factor $Z(\phi)$. 

For simplicity we wish to consider isotropic solutions in the bulk, and hence we will assume a background for the metric and gauge field of the form
\be
ds^2 = - U dt^2 + U^{-1} dr^2 + e^{2V}(dx^2 + dy^2) \nonumber
\ee
\be
A = a(r) dt \nonumber
\ee
We assume that the geometry has a regular horizon at $r=r_+$ where the gauge field vanishes $a \sim (r - r_+)$ and $U \sim 4 \pi T (r - r_+)$. As the radial coordinate $r \rightarrow \infty$ we assume that the metric approaches anti-de Sitter space and that the gauge field approaches a constant which is interpreted as the chemical potential, $\mu$, in the boundary.

Associated to the chemical potential is a constant charge density, ${\cal Q}$, which is identified with the conserved electric flux of the bulk theory ${\cal Q} = -e^{2 V} Z(\phi) a'$. In order to have a finite conductivity at ${\cal Q} \neq 0$ we must break translational invariance. This is done by demanding that the scalar fields are non-vanishing on the boundary $\chi_1 \rightarrow k x$, $\chi_2 \rightarrow k y$ as $r \rightarrow \infty$. This corresponds to introducing oscillatory lattices in the scalar fields $\phi_i$. 

\paragraph{DC conductivity}

Before proceeding to calculate the Hall conductivity, it will be important to first highlight some features of the DC conductivity of these holographic models. There has recently been a large amount of progress in obtaining analytic expressions for the transport properties of holographic theories  \cite{univdc, lattices, aristosdc, blaise, withers, thermo}. The key idea \cite{univdc} is that the DC conductivity does not evolve in the radial direction and hence can be expressed solely in terms of horizon data. In particular, for the above holographic models, the resulting expression derived in \cite{aristosdc} is 
\be
\sigma_{DC} = \bigg [Z(\phi) + \frac{4 \pi {\cal Q}^2 }{ k^2 \Phi(\phi) s} \bigg ]_{r_+}
\label{dccond}
\ee
where $s = e^{2V}/4\pi|_{r_+}$ is the entropy density. 

\begin{figure}
\begin{center}
\resizebox{50mm}{!}{\includegraphics{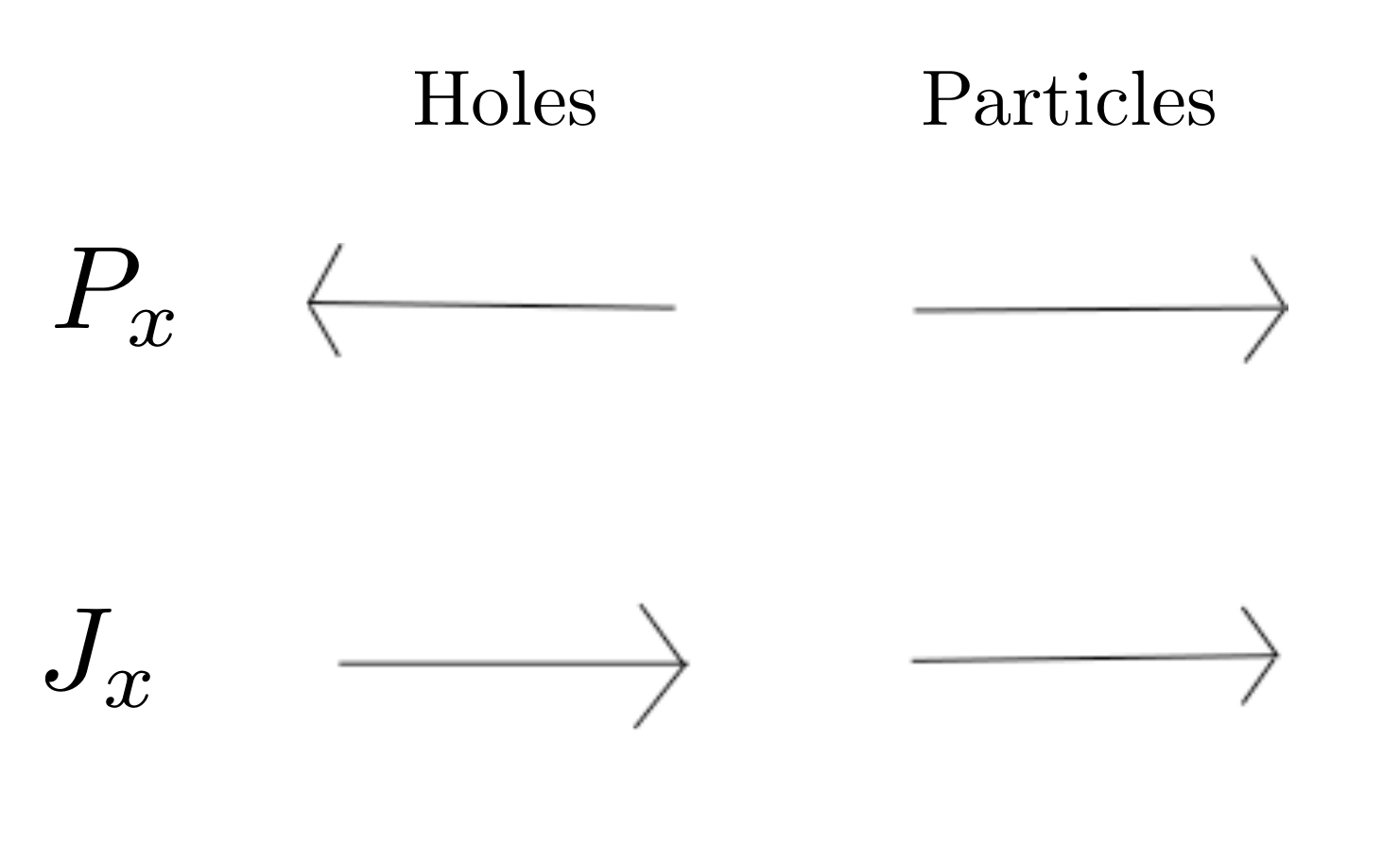}}
\caption{At weak coupling, the conductivity of a charge-conjugation symmetric theory can be understood as arising from particle hole pairs of opposite momenta.}
\label{dccurrent}
\end{center}
\end{figure}

An important observation is the division of the conductivity into two distinct terms. A precise distinction can be made by comparison to the electrothermal conductivity, $\bar{\alpha}$, computed in \cite{thermo} 
\be
\bar{\alpha} = \bigg[ \frac{4 \pi {\cal Q}}{k^2 \Phi(\phi))} \bigg]_{r_+}
\label{thermo}
\ee
from which we can see that the first term in the DC conductivity, $Z(\phi)|_{r_+}$, does not contribute to the electrothermal conductivity. Such a term is already present at ${\cal Q}=0$, where the theory is charge conjugation symmetric. In a weakly coupled system one can understand this `charge conjugation symmetric' conductivity as arising from particle hole pairs moving in opposite directions, as illustrated in Fig.~\ref{dccurrent}. However, we stress that for the holographic theories discussed here, which are strongly interacting and contain no quasiparticles, this intuition is suggestive at best. Nevertheless, since $\sigma_{DC}$ is even under charge conjugation symmetry, there is a non-zero conductivity for ${\cal Q} =0$ even at strong coupling.

A more surprising and novel feature of these holographic theories is that this  `charge-conjugation symmetric' conductivity  $\sigma_{ccs} = Z(\phi)|_{r_+}$ remains present even at finite density. For the case of relativistic free fermions, for instance in graphene, the addition of a chemical potential would introduce a gap for particle-hole creation proportional to $\mu$ and we would expect such a term to be exponentially suppressed below this scale. In contrast, for the strongly coupled holographic theories discussed here, $\sigma_{ccs}$ can have a power-law dependence on $T$ even at finite density \cite{blaise2010}. 

When  ${\cal Q} \neq 0$ we also have to consider the second term in \eqn{dccond}. For a translationally invariant theory this term would diverge, but is rendered finite in our models by the presence of the lattice. As we outlined in the introduction we will refer to this term, which is associated with momentum dissipation, as $\sigma_{diss}$. The key point of \eqn{dccond} is that the finite density conductivity consists of two terms added together - that is they follow an `inverse-Matthiessen' law
\be
\sigma_{DC} = \sigma_{ccs} + \sigma_{diss}
\label{mathiessen}
\ee
%
%
%
%
In particular we reiterate that within these holographic models both of these terms can remain at low energies, even in the presence of a chemical potential. 

\paragraph{Hall conductivity} Having explained the salient features of the DC conductivity, we now wish to generalise the techniques of \cite{aristosdc} to calculate the Hall conductivity. In Lorentz invariant theories, 
the conductivity in the presence of a magnetic field is constrained to obey the simple form $\sigma_{xx} = 0$, $\sigma_{xy} = {\cal Q}/B$ which was originally reproduced from holography in \cite{hall}. 

Calculating the Hall conductivity in theories without translational invariance is more complicated. To do this, we consider the same class of models as before, but we add a magnetic field $A_{y} = B {x}$ to the background. In order to calculate the conductivity, we perturb the background solution by a constant electric field $A_x = - E_x t$ as performed in \cite{aristosdc}. The bulk equations then force us to turn on other fields, for which a consistent ansatz is
\begin{eqnarray}
A_{x_i} &=& - E_i t + \delta a_{x_i} (r) \nonumber \\
g_{tx_i} &=& e^{2V} \delta h_{tx_i} (r) \nonumber \\
g_{rx_i} &=& e^{2V} \delta h_{rx_i}(r) \nonumber \\
\chi_i &=& kx_i + \delta \chi_i(r) \nonumber 
\end{eqnarray}
where $i$ runs over $(1,2)$ and we of course mean that $x_1 = x, x_2 = y$.  
As is well-known in these calculations, the trick is to find quantities that are independent of the bulk radial coordinate. These are provided for us by the perturbed Maxwell equations, which give us the two constant fluxes
\begin{eqnarray}
J_{x} &=& - Z(\phi) U \delta a_{x}' + {\cal Q } \delta h_{tx} -  B Z(\phi) U \delta h_{r y}  \nonumber \\
J_{y} &=& - Z(\phi) U \delta a_{y}' + {\cal Q}\delta h_{ty} + B  Z(\phi) U \delta h_{r x} 
\end{eqnarray}
from which we can evaluate the conductivities via the ratios
\begin{eqnarray}
\sigma_{x x} = \frac{ J_{x}}{E_x} \;\;\;\;\;\; \sigma_{x y} = \frac{ J_{y}}{E_x} \nonumber 
\end{eqnarray}
Since $J_x$ and $J_y$ are constants, we can calculate these ratios anywhere in the bulk. The simplest place to do this is at the horizon, where the constraints of regularity are enough to determine the conductivity. That is, we demand the smooth behaviour
\begin{eqnarray}
\delta a_{x_i} &=&  -\frac{E_i}{4 \pi T} \mathrm{ln}( r - r+) + {\cal O}(r - r_+) \nonumber \\
\delta \chi_i &=& {\cal O}((r - r_+)^{0}) \nonumber \\
\delta h_{tx_i} &=& U \delta h_{rx_i} + {\cal O}(r - r_+)
\end{eqnarray}
The quickest way to evaluate the conductivity is to plug these requirements into the $t-x$ component of Einstein's equations. This results in a pair of simultaneous equations for the value of $\delta h_{tx_i}$ at the horizon. 
\begin{eqnarray}
(B^2Z(\phi) + e^{2V}k^2 \Phi(\phi)) \delta h_{t x} &-&  B Z(\phi) e^{2V} a' \delta h_{ty} \nonumber \\
&=& - e^{2V} Z(\phi) a' E_x \nonumber \\
(B^2Z(\phi) + e^{2V}k^2 \Phi(\phi)) \delta h_{t y} &+& B Z(\phi) e^{2V} a' \delta h_{tx} \nonumber \\ 
&=& B Z(\phi) E_x \nonumber
\end{eqnarray}
%
%
%
Inverting these equations gives the values of the graviton at the horizon, from which we can proceed to extract the Hall conductivity
\begin{eqnarray}
\sigma_{x x} &=& \frac{e^{2 V} k^2 \Phi (B^2 Z^2 + {\cal Q}^2 + Z e^{2 V} k^2 \Phi)}{(B^2Z + e^{2V} k^2 \Phi)^2 + B^2 {\cal Q}^2} \bigg|_{r_+} \nonumber \\
\sigma_{x y} &=&  \frac{B {\cal Q} (B^2Z^2 + {\cal Q}^2 + 2 Z e^{2 V} k^2 \Phi) }{(B^2Z + e^{2V} k^2 \Phi)^2+ B^2 {\cal Q}^2} \bigg|_{r_+} 
\label{hallconductivity}
\end{eqnarray}
%
%


 We are now able to turn to the question of ultimate interest, which is to calculate the Hall angle $\theta_H = \sigma_{x y}/\sigma_{x x}$
 for holographic theories in a magnetic field. Whilst transport properties in a magnetic field can be unfamiliar, the Hall angle is especially simple. In many ways it behaves like the familiar DC conductivity- in the absence of a lattice it is an infinite delta function, that will now be resolved via momentum dissipation into a Drude peak. 
 
The holographic results above imply that the Hall angle takes the somewhat clumsy form
\begin{equation}
\theta_H = \frac{B {\cal Q}}{e^{2V} k^2 \Phi}\bigg[ \frac{ B^2 Z^2 + {\cal Q}^2 + 2 Z e^{2 V} k^2 \Phi}{ B^2 Z^2 + {\cal Q}^2 + Z e^{2 V} k^2  \Phi}\bigg] \bigg|_{r_+}
\end{equation}
Whilst this formula is complicated, we can extract the physics by noticing that for all geometries the quantity in square brackets is simply a number bounded between one and two.
We can therefore deduce that the scaling of the Hall angle is predominantly controlled by the overall factor outside the brackets which may be written as
\begin{equation}
\theta_H \sim \frac{ 4 \pi B {\cal Q} }{k^2 \Phi(\phi) s} \bigg|_{r_+}
\label{hallangle}
\end{equation}
Furthermore, for small magnetic fields the thermodynamic and lattice factors appearing in \eqn{dccond} and \eqn{hallangle} must agree and so we may write
\begin{equation}
\theta_H  \sim \frac{B}{{\cal Q}} \sigma_{diss}
\label{diss}
 \end{equation}
The central point of this letter is the observation that, in contrast to the DC conductivity, there is only a single contribution to the Hall angle. 

\begin{figure}
\begin{center}
\resizebox{50mm}{!}{\includegraphics{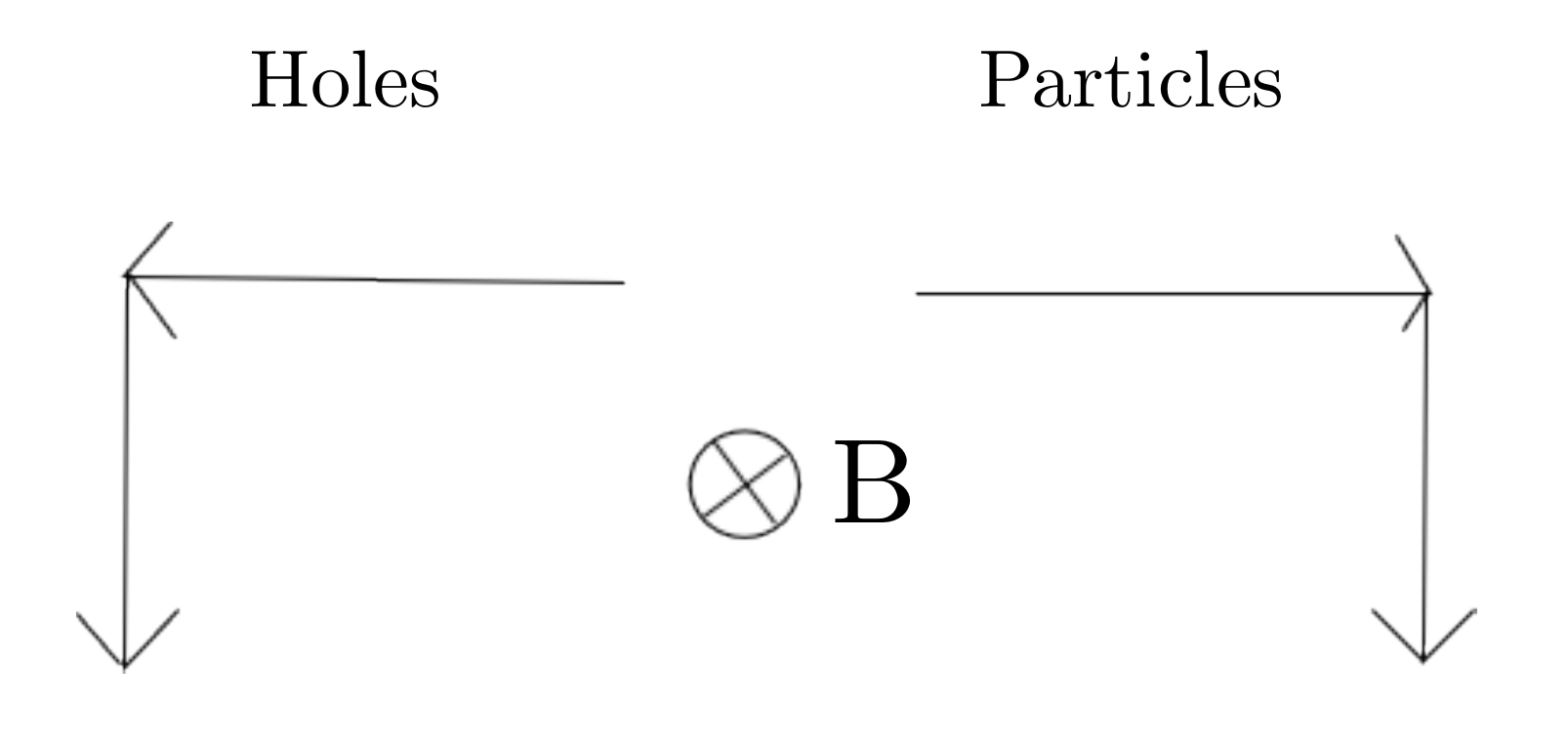}}
\caption{In the presence of a magnetic field, the particle-hole pairs responsible for $\sigma_{ccs}$ are deflected in the same direction. They therefore cannot carry a Hall current.}
\label{hallcurrent}
\end{center}
\end{figure}

In particular, there is no additive contribution to the Hall angle analogous to the `charge-conjugation symmetric' conductivity in $\sigma_{DC}$. In fact, at least at weak coupling, it is easy to  see why this should be the case. Recall that this current was carried by particle-hole pairs moving in opposite directions. Upon adding a magnetic field, these pairs are deflected in the same direction and hence they do not contribute to the Hall conductivity $\sigma_{xy}$ (Fig.~\ref{hallcurrent}). This simple observation continues to hold in strongly coupled theories - $\sigma_{xy}$ is odd under charge conjugation symmetry and hence must vanish when ${\cal Q} = 0$.

%
%

Motivated by the experimental results, our goal is to understand how we can obtain different scalings in the Hall angle and DC conductivity. It is easy to reproduce the original puzzle of the Hall angle. For geometries where the lattice is very small, $\sigma_{diss} \gg 1$, then the DC conductivity is dominated by the second term in \eqn{dccond}, and so scales in the same manner as the Hall angle. 

This result should not be a surprise. In this regime,  the correct framework to describe strongly coupled transport is the memory matrix \cite{hydro, impure, sandiego}. Within this framework, every operator that has a projection onto the momentum operator, such as the electric and Hall currents, is controlled by the momentum relaxation rate. The physics is dominated by this single timescale and hence the Hall angle and DC conductivity must agree. 

However, a simple resolution of this puzzle is equally clear. Outside of this momentum dissipation dominated regime, we must also consider the `charge-conjugation symmetric' contribution to $\sigma_{DC}$. For strong momentum dissipation this dominates the DC conductivity, $\sigma_{DC} \sim \sigma_{ccs}$, and so within the holographic models this is given by the dilaton factor. Conversely, the Hall angle is still given by $\theta_H \sim \sigma_{diss}$, and so is determined by the profile of the lattice according to \eqn{hallangle}. Whilst in this letter we do not consider any specific model, there is not generically a simple relationship between the dilaton factor and the lattice profile. As such we can deduce that general holographic models will have very different scalings in the Hall angle as to the DC conductivity. 

\paragraph{Discussion}

We have shown that, within a large class of holographic lattice models, one will generically find different scalings in the Hall angle and the DC conductivity. Ultimately this was made possible by the presence of a `charge conjugation symmetric' contribution to the DC conductivity that was absent from the Hall angle. 
We emphasised that at weak coupling one would expect such a term to be exponentially suppressed. The novel step in our argument is the realisation that holography provides examples of strongly interacting systems where this is no longer the case. Hence, despite charge-conjugation symmetry being badly broken by the chemical potential, one can find a regime, of strong momentum dissipation, where $\sigma_{ccs}$ dominates the low energy conductivity. Since the Hall angle is independent of $\sigma_{ccs}$, it is then simple to obtain anomalous scaling. 

The existence of $\sigma_{ccs}$, and hence this method of obtaining an anomalous Hall scaling, is more general than these specific lattice models. For instance, in \cite{hydro} the transport properties of Lorentz invariant CFTs deformed by a net charge density, ${\cal Q}$, and magnetic field $B$ were studied using hydrodynamics. The DC and Hall conductivities calculated in \cite{hydro} take the same form as the holographic lattice results \eqn{dccond} and \eqn{hallconductivity} \footnote{To make this comparison one identifies the $\sigma_Q$ of \cite{hydro} with our $\sigma_{ccs}$ and the momentum relaxation rate of hydrodynamics as $\tau^{-1} = e^{2 V} k^2 \Phi|_{r_+} / ( {\cal E} + {\cal P})$. Here ${\cal E}$ and ${\cal P}$ are the energy density and pressure respectively.}. Although it was not appreciated at the time, our general arguments are equally applicable to these models and it is possible to describe an anomalous Hall angle within relativistic hydrodynamics. 

More recently, an anomalous scaling of the Hall angle was identified in probe brane models of holographic transport \cite{karch}. Although the conductivity of these models takes a more complicated form than \eqn{mathiessen}, it is still possible to identify a `charge-conjugation symmetric' contribution to $\sigma_{DC}$. The mechanism introduced here can be used to provide a microscopic understanding of these results. 

Given the general nature of these ideas, we end this letter with the a brief discussion of the implications of our work for the experimental phenomenology of the cuprates. Recall that we have two contributions to the conductivity, a `charge-conjugation symmetric' term $\sigma_{ccs}$ and another from explicit charge density relaxed by some momentum dissipation, $\sigma_{diss}$. These are additive in the DC conductivity following an inverse Mathiessen rule \cite{criticallook}
\be
\sigma_{DC} = \sigma_{ccs} + \sigma_{diss}
\ee
Conversely the temperature scaling of the Hall angle is always
\be
\theta_H \sim \frac{B}{\mathcal{Q}}\sigma_{diss}
\ee
In order to obtain an anomalous Hall scaling we needed the DC conductivity to be dominated by the `charge-conjugation symmetric' term. Therefore the linear resistivity of the cuprates would arise from a scaling - $\sigma_{ccs} \sim 1/T$, whilst we could reproduce the scaling of the Hall angle provided we have that $\sigma_{diss} \sim 1/T^2$. This should be contrasted with the many attempts to use holography to obtain a linear resistivity which assume that $\sigma_{diss}$ dominates the conductivity \cite{pauli, janrichard}. Our results suggest it would be difficult to accommodate the Hall angle anomaly within this approach. Rather, our ideas are more inline with the recent suggestion that a linear resistivity could arise from universal incoherent transport \cite{seanincoh}. 


Having combined these two contributions, the resistivity, $\rho$, predicted by this approach would take the general form
\be
\rho \sim T^2/(W + T)
\ee
where $W$ is a model dependent energy scale. In our picture, the `charge conjugation symmetric' conductivity dominates for $T \gg W$ and we are left with the linear resistivity. Conversely for $T \ll W$, momentum relaxation is very weak and the resistivity passes back over to a Fermi-liquid like $T^2$ law. Although arising from a completely different model, a similar form for the resistivity was recently presented in \cite{hiddenfl, hiddenflprl}. In particular, it was emphasised that this crossover is remarkably similar to the behaviour of the overdoped cuprates and was found in \cite{hiddenflprl} to provide a good fit to experimental data. Further, the existence of both $T$ and $T^2$ contributions to the resistivity has also been seen in \cite{hussey} - albeit whilst fitted to a more conventional Mathiessen law. 

Finally we comment that the dichotomy in scalings will be evident in more general transport properties. In particular, the thermoelectric current is also observed to show the same anomalous 
behaviour as the Hall angle \cite{Coleman}. It is clear that our picture reproduces this result - as is explicit in \eqn{thermo}, the thermoelectric conductivity only receives a contribution from net charge and so, like the Hall angle, is proportional to $\sigma_{diss}$. 

Whilst it is hard to make any concrete comparisons with experiment due to the qualitative nature of this work, it is encouraging that these simple features hold up. It would of course be of great interest to develop specific models realising this scenario, so that a more detailed comparison with experimental phenomenology can be performed.

Our thanks to Richard Davison and David Tong for many useful conversations, to Jerome Gauntlett for collaboration with one of us on related work, and to Sean Hartnoll for comments on the manuscript.  We are also grateful to the two anonymous referees whose feedback greatly improved the presentation of these ideas. MB and AD are supported by STFC and by the European
Research Council under the European Union's Seventh Framework Programme
(FP7/2007-2013), ERC Grant agreement STG 279943, ``Strongly Coupled Systems''. MB thanks Girton College for their continued support over the last seven years. This work was revised whilst MB was a Junior Research Fellow at Churchill College, Cambridge.

\end{document}